Magnetic field processing to enhance critical current densities of $MgB_2$ superconductors

S.X. Dou[1], W.K. Yeoh[1], O. Shcherbakova[1], J. Horvat[1], M.J. Qin[1], Y. Li[2], Z.M. Ren[2], P. Munroe[3]

[1]Institute for Superconducting and Electronic Materials, University of Wollongong, Northfields Ave. Wollongong, NSW 2522 Australia
[2]Department of Materials Science and Engineering, Shanghai University, 149 Yanchang Rd. Shanghai 200072 P.R. China
[3]Electron Microscope Unit, University of New South Wales, Sydney, NSW 2000, Australia

**Abstract**

Magnetic field of up to 12 T was applied during the sintering process of pure $MgB_2$ and carbon nanotube (CNT) doped $MgB_2$ wires. We have demonstrated that magnetic field processing results in grain refinement, homogeneity and significant enhancement in $J_c(H)$ and $H_{irr}$. The $J_c$ of pure $MgB_2$ wire increased by up to a factor of 3 to 4 and CNT doped $MgB_2$ by up to an order of magnitude in high field region respectively, compared to that of the non-field processed samples. $H_{irr}$ for CNT doped sample reached 7.7 T at 20 K. Magnetic field processing reduces the resistivity in CNT doped $MgB_2$, straightens the entangled CNT and improves the adherence between CNTs and $MgB_2$ matrix. No crystalline alignment of $MgB_2$ was observed. This method can be easily scalable for a continuous production and represents a new milestone in the development of $MgB_2$ superconductors and related systems.

The new superconductor, $MgB_2$, has made a significant impact on the research and development of superconductors since its discovery[1]. The special feature of the two-gap superconductivity[2] and lack of week links at the grain boundaries[3] makes $MgB_2$ highly tolerant for doping which has been successfully used to enhance the critical current density, $J_c$ and the upper critical field, $H_{c2}$[4-8]. Carbon and silicon carbide doping resulted in a significant increase of in-field $J_c$ and $H_{c2}$, and these records still stand for $MgB_2$[5-11]. To further advance the development of $MgB_2$ for applications we report a new method of combining the advantages of magnetic field processing and of doping for processing $MgB_2$ superconductors. Magnetic field processing technology has been proved to be a powerful tool to produce aligned CNT in composites and neat macroscopic membranes[12-14] and control the phase transformation and behavior of the melts during condensation processes, resulting in major improvements in material properties[15,16]. Magnetic field processing has also been used to achieve the desired texture and improved $J_c$ performance in HTS[17-20]. In processing of $MgB_2$ bulk and wires the reaction *in-situ* technique in combination with the powder-in-tube (PIT) method has been used to produce the wires with the best field performance[21-23]. Other advantages of this process include easy fabrication of coils and the ability to incorporate dopants and additives, which are important for improvement of flux pinning and $H_{c2}$. In the *in-situ* reaction process, Mg melts before the $MgB_2$ formation by solid state reaction, provided the heating rate is high enough. The presence of a liquid phase provides a window of opportunity for applying a magnetic field processing technique to achieve a crystalline refinement, homogeneous distribution of additives and inclusions and possible alignment of both matrix materials and additives.

In this work, a standard powder-in-tube method was used for Fe clad $MgB_2$ wire[23]. Powders of magnesium (99%) and amorphous boron (99%) were well mixed with 0 and 10 wt% of multi-wall CNTs (OD: 20 nm and length: 0.5–2 µm), and thoroughly ground. The Fe tube had an outside diameter (OD) of 10 mm, a wall thickness of 1 mm, and was 10 cm long with one end of the tube sealed. The mixed powder was filled into the tube and the remaining end was blocked using aluminum bar. The composite was drawn to a 1.4 mm diameter wire. Several short samples 2-3 cm in length were cut from the wire. These pieces were sealed in Fe tube and then sintered in a tube furnace at 800 °C for 30min to 3 hours in a 4 and 5 T pulsed, and a 10 T and 12 T static magnetic field ($H_a$) with a high heating rate at 20 °C/min, and finally furnace-cooled to room temperature. The processing field was applied parallel to the wire axis before temperature ramped up. The fast heating is to ensure the melting of Mg before solid reaction between Mg and B. The same process was repeated without field to make standard samples for comparison.

The magnetization of cores taken from the wires was measured at 5 K and 20 K using a Quantum Design Physical Property Measurement System (PPMS,) with a magnetic field sweep rate of 50 Oe/s and amplitude up to 8.5 T. Since there is a large sample size effect on the magnetic $J_c$ for $MgB_2$ fabricated with reaction in-situ process[24] all the samples for measurement were made to the same size for comparison. The magnetic $J_c$ was derived from the height of the magnetization loop using critical state model. Samples were measured with the field, $H$, applied perpendicular and parallel to the wire axis. The transport $J_c$ was measured with the four-probe method using a pulsed current source for critical currents down to about 70A. In high field region, providing critical currents lower than 1A, a DC current source was used to measure the transport $J_c$.

Fig. 1 shows the magnetic $J_c$ for the undoped $MgB_2$ with the measurement field, $H$, applied perpendicular to the axis of wire core. The processing field, $H_a$, was applied parallel to the wire axis during sintering process. The field processed samples show enhancement in $J_c(H)$ in all the field and temperature range. In particular, there is a more pronounced increase of $J_c$ in higher field region. The $J_c$ for the sample processed in 5 T field increases by up to a factor of 4 at 5 K and 7 T and a factor of 3 at 20 K and 4 T, compared with the standard undoped $MgB_2$ sample without field processing. There is no anisotropy in $J_c$ : $J_c$ is the same for parallel and perpendicular measurement field $H$, while the processing field $H_a$ was parallel to the wire axis. This indicates that there is no preferred crystalline orientation of $MgB_2$ due to the magnetic field processing.

Fig. 2 shows the magnetic $J_c(H)$ curves for the 10% CNT doped $MgB_2$ samples processed without field and with 5 T pulsed and 10 T static field. 5 T pulsed magnetic field processing results in better $J_c$ in the entire field region while more pronounced $J_c$ enhancement is observed in the high field region using a 10 T field processing. The extent of the enhancement in $J_c$ due to the field processing increases with increasing measurement field $H$, ranging from a factor of 1.5 in low fields up to one order of magnitude in high fields. For example, the $J_c$ in $H \perp H_a$ increases by a factor of 4 and 8 at 5 K and 8 T for the sample processed in the 5 T and the 10 T field respectively, compared to the sample without field processing. The inset in figure 2 shows anisotropy of $J_c(H)$ in $H \perp H_a$ and $H // H_a$. As the $MgB_2$ crystals are not aligned in the magnetic field, the anisotropy in CNT doped $MgB_2$ may be attributable to the alignment of CNT in the field.

Fig. 3 shows a comparison of the transport $J_c(H)$ at 20 K for CNT doped $MgB_2$ wire processed without applied field and with 10 T static field at 800 °C for 3 h. The results show the same trend as magnetic $J_c(H)$, the improvement in $J_c$ increases with increasing measurement field. It is worthy noting that the irreversibility field ($H_{irr}$) for 10 T processed CNT doped $MgB_2$ wire reached 7.7 T (using the usual 100 A/cm$^2$ criterion), which is as good as for the best value for the SiC doped $MgB_2$[5]. The inset is the comparison of normalized resistance for the CNT doped samples processed at 800°C for 30 min in 0 T and 12 T field. The results indicate that field processing reduces the resistivity of $MgB_2$ sample by about 30%.

To understand the mechanism behind the enhancement in $J_c(H)$ one must consider the behavior and properties of $MgB_2$ and major inclusions during magnetic field processing. The magnetic processing has several beneficial effects, including refinement of crystallites, homogenization of inclusions and alignment of parent or additive materials. To achieve crystalline alignment in a magnetic field there are two requirements: liquid formation and crystallites that have a size sufficient to give them an anisotropy energy larger than the energy associated with the thermal disordering effect (~ $kT$). The anisotropy energy is proportional to the volume of each crystallite ($V$), to the anisotropy of the paramagnetic susceptibility ($\Delta\chi$) and the square of the applied magnetic field ($H_a^2$). Studies of the magnetism of the normal state of MgB$_2$ show that the net susceptibility will be very small, of the order of 10$^{-6}$ emu/mol[25]. Furthermore, the size of each of the crystallites of MgB$_2$ is very small (about 100nm to 500nm). These factors may account for the nonoccurrence of crystalline orientation in the MgB$_2$ processed in magnetic field. The improvement of $J_c(H)$ in high field region for the field processed MgB$_2$ is attributable to the grain refinement as shown by the SEM images in Fig 4. It is noted that microstructure for the 5 T field processed MgB$_2$ (Fig. 4(b)) is much more homogeneous than the non-field processed sample (Fig. 4 (a)). Also the grain size for the former is bellow 200 nm while the latter has a wide range of size up to more than 1000 nm. As smaller grains will have stronger grain boundary pinning than the larger ones this may account for the improvement in $J_c(H)$ in higher field region.

As reported in previous work the added CNTs were entangled in the MgB$_2$ matrix for the non-field processed sample and appeared as mainly bare CNTs and not well bonded to MgB$_2$ matrix[8]. In contrast, the CNTs in magnetic field processed sample are straightened and imbedded in MgB$_2$ matrix as shown in Fig. 5(a). Fig. 5(b) shows focused ion beam (FIB) assisted scanning electron micrographs (SEM) for the CNT doped MgB$_2$ processed with a 5 T pulsed magnetic field applied along the wire axis direction. The magnetic field processing converts the CNT agglomerates to whiskers with length of about 2-5μm and diameter of 200-400 nm which is ten times thicker than the diameter of CNTs. These whiskers are formed with CNTs as nucleation centers for the MgB$_2$ growth in the area of CNTs agglomerates under magnetic field. These whiskers are mostly straightened with good contact among them. These MgB$_2$ whiskers would be good conductors, accounting for the enhancement of $J_c$ in low field region. Fig. 5 (c) is a TEM image showing a number of aligned and straight CNTs imbedded in MgB$_2$. The aligned CNTs will improve the grain connectivity and act as efficient flux pinning sites. This observation is consistent with the observed anisotropy in $J_c(H)$ for field processed CNT doped MgB$_2$, which is lacking for pure MgB$_2$. Ajiki et al[26], Lu[27] and Ramirez et al.[28] found that the susceptibility, χ, of CNT is large enough to be aligned in certain magnetic field.

The advantages of the magnetic field processing include the provision of a unique window of opportunity to control the microstructure and the behavior of various additives that have a response to magnetic field. We can extend the conventional materials processing variables of pressure, composition and temperature, f(P,C.T) to include field (H) and magnetization (M) of the matrix and additives, f(P,C,T,H,M). For the undoped MgB$_2$, magnetic field processing can improve the $J_c$ in the entire field region without need of any doping. For CNT doped MgB$_2$, magnetic field processing enhances $J_c(H)$ in all the field and temperature range by up to an order of magnitude. Despite the unoptimized processing conditions used, the significant benefit of combining the advantages of magnetic field processing and element doping is clearly demonstrated. We can manipulate various combination of the parameters such as field strength, processing temperature and time, and properties of additives to achieve an optimal enhancement in $J_c(H)$. This opens up an interesting research field so that all the additives to MgB$_2$ and related systems studied thus far must be revisited under magnetic field processing conditions and a better enhancement in $J_c$ is expected. As the formation of MgB$_2$ in the reaction in-situ technique is a rapid process (10 to 30 min.), magnetic field processing can be easily used in large-scale production.

**Acknowledgements**  The work was supported by the Australian Research Council, Hyper Tech Research Inc, OH, USA, and Alphatech International Ltd, NZ. We thank Drs X.L. Wang, A. Pan, T. Silver, and Mr. M. Tomsic for helpful comments. W. K. Yeoh received an Australia-Asia Award funded by the Australian Government. Z.M. Ren and Y. Li would like to thank the Natural Science Foundation of China for their support under grants (No. 59871026, 50225416, 50234020).

**Correspondence** and request should be addressed to S.X. Dou (shi_dou@uow.edu.au)

Figure captions

Fig. 1 Critical current density as a function of applied magnetic field for the pure $MgB_2$ processed at 800°C for 30 min without magnetic field, and with 5 T static field

Fig. 2 $J_c$ vs H for CNT doped $MgB_2$, processed at 800°C for 30 min without magnetic field, and with 5 T pulse and 10 T static field respectively. The inset shows anisotropy of $J_c$ at 5 K for CNT doped $MgB_2$ processed at 10 T field with processing $H_a$ applied parallel to and perpendicular to the wire sample axis

Fig. 3 The transport $J_c$ vs H for CNT doped $MgB_2$ with measurement field perpendicular to the wire axis. The inset is a comparison of the normalized resistivity for the CNT doped $MgB_2$ wire processed in 0 T and 12 T field

Fig. 2 A comparison of SEM images for pure $MgB_2$, processed at 800°C for 30 min without magnetic field (a) and with 5 T pulse field (b).

Fig. 5 Transmission electron micrographs for CNT doped $MgB_2$ processed with 5 T pulsed field (a), FIB-SEM micrographs of straight $MgB_2$ whisks formed on the CNTs (b) and TEM image of aligned CNTs imbedded in $MgB_2$ matrix (c).

Fig. 1

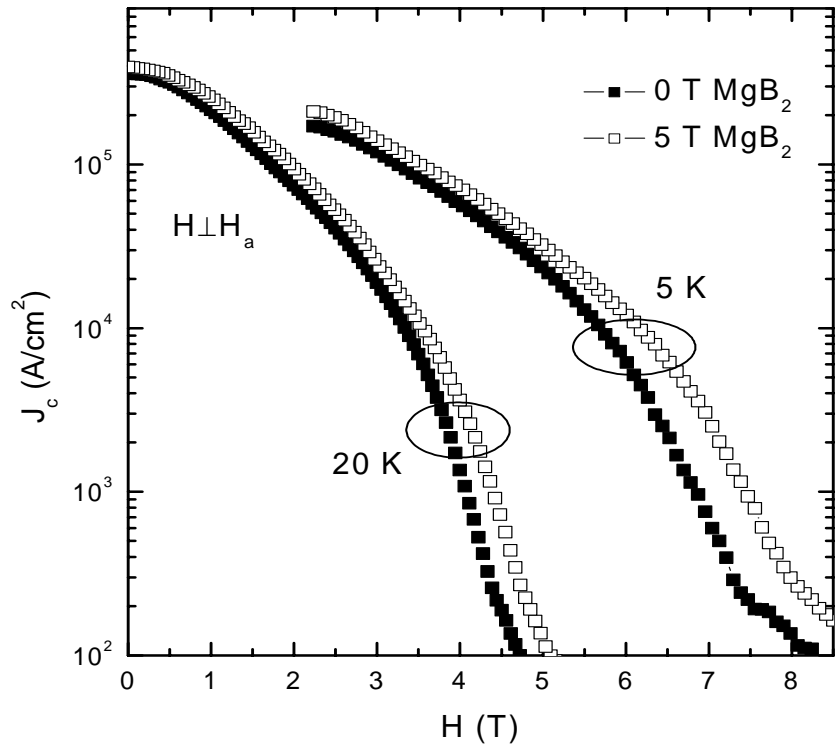

Fig. 2

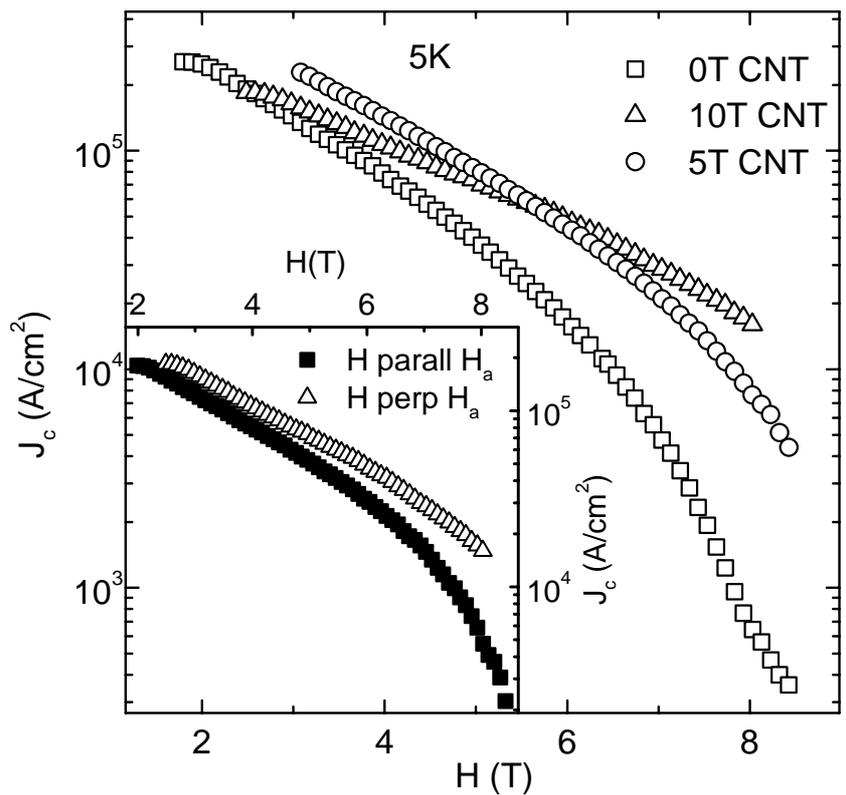

Fig 3

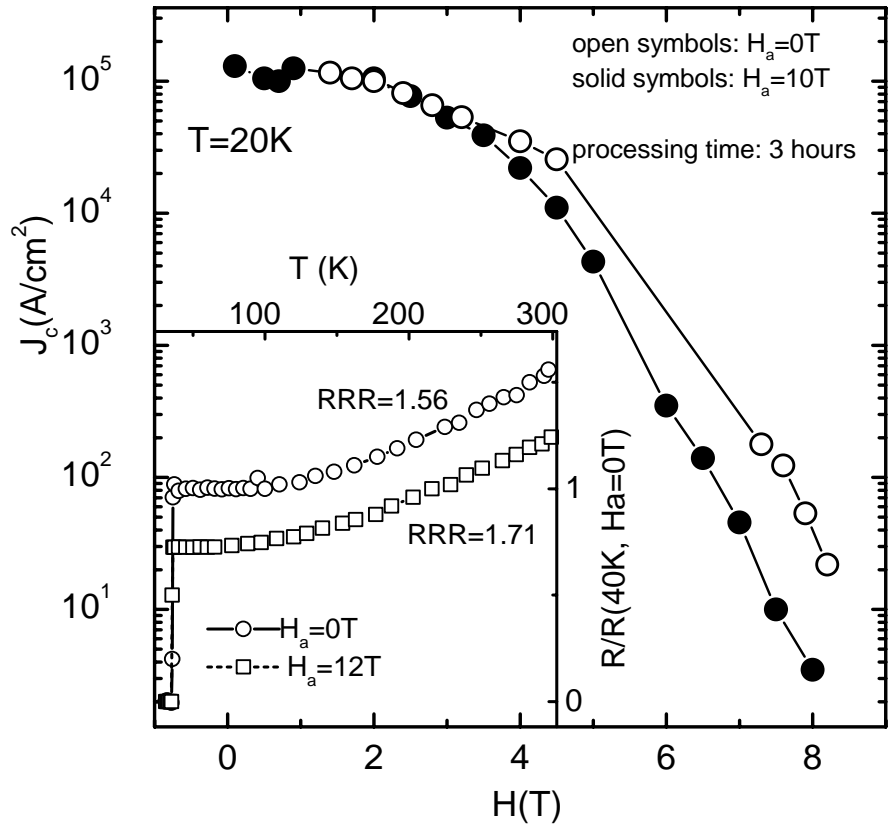

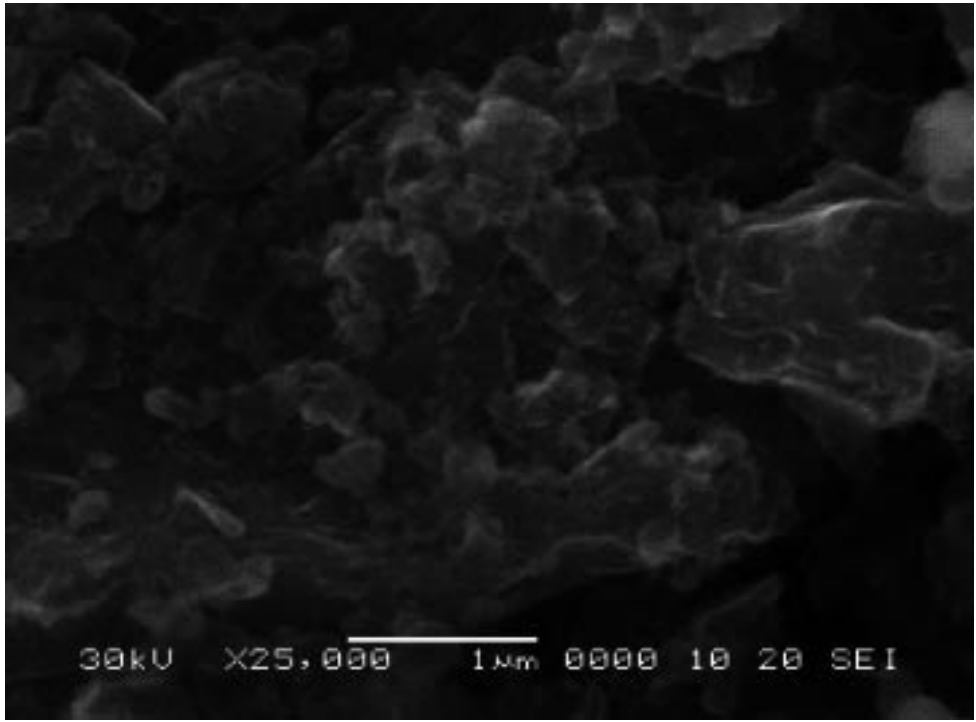

Fig. 4(a)

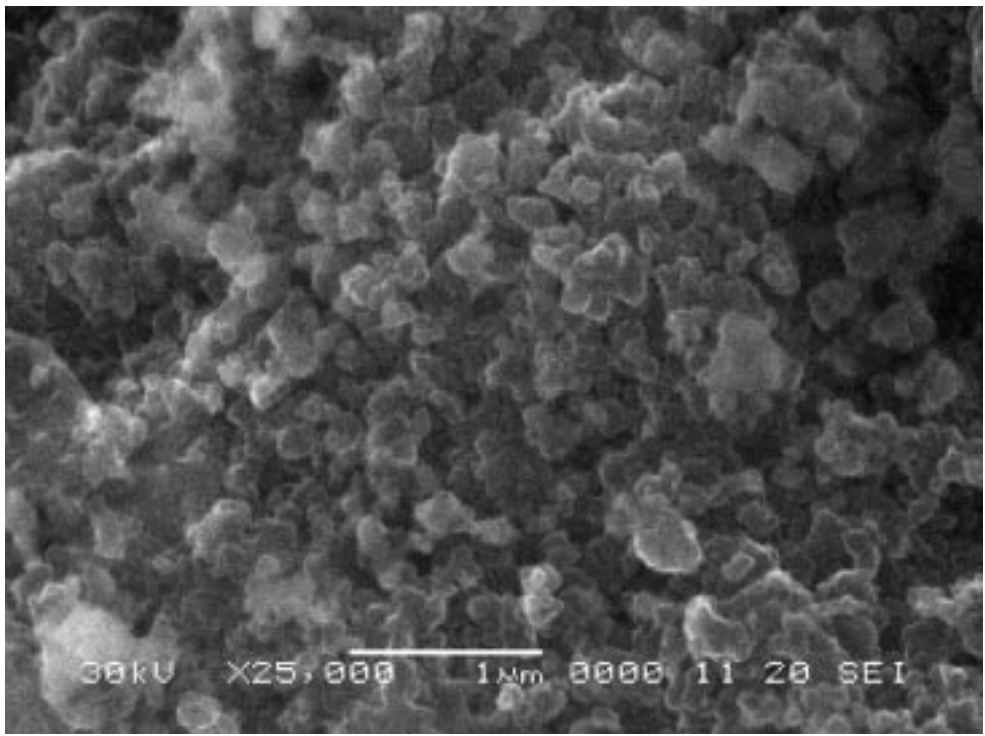

Fig. 4(b)

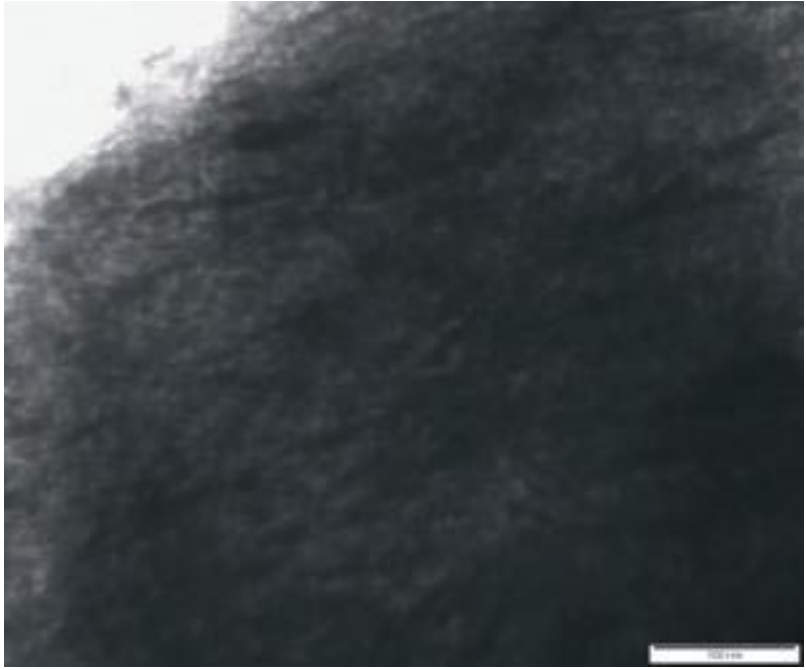
Fig 5 (a)

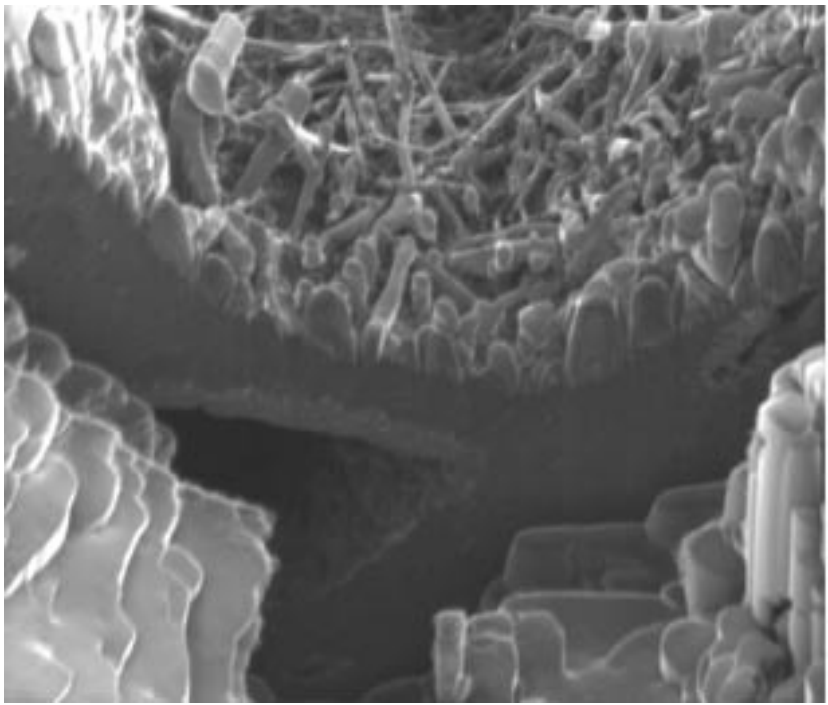
Fig. 5 (b)

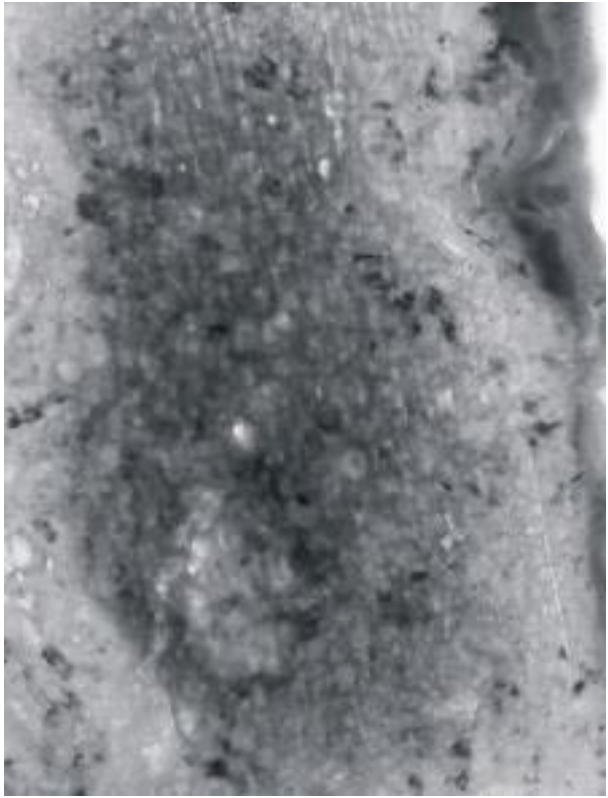

Fig. 5(c)